\documentclass[10pt,twocolumn,a4paper]{IEEEtran}
\usepackage{amsmath}
\usepackage{graphicx}

\newcommand{\figuragrossa}[3]
{
\begin{figure}
  \centering
 \includegraphics[width=7cm]{#1}
  \caption{#2}\label{#3}
\end{figure}
}

\newcommand{\figura}[3]
{
\begin{figure}
  \centering
 \includegraphics[width=8cm]{#1}
  \caption{#2}\label{#3}
\end{figure}
}
\title{On the Behavior of the Distributed Coordination
Function of IEEE 802.11 with Multirate Capability under General
Transmission Conditions}
\author{\authorblockN{F. Daneshgaran, M. Laddomada, F. Mesiti, and M.
Mondin\thanks{This work has been supported by Euroconcepts, S.r.l.
(http://www.euroconcepts.it)}
\thanks{F. Daneshgaran is with ECE Dept., California State University,
Los Angeles, USA.}
\thanks{M. Laddomada, F. Mesiti, and M. Mondin are with DELEN, Politecnico di Torino, Italy.}}
\authorblockA{}}
\begin{document}
\maketitle
\begin{abstract}
The aim of this paper is threefold. First, it presents a
multi-dimensional Markovian state transition model characterizing
the behavior of the IEEE 802.11 protocol at the Medium Access Control layer which accounts for
packet transmission failures due to channel errors modeling
both saturated and non-saturated traffic conditions. Second, it
provides a throughput analysis of the IEEE 802.11 protocol at the
data link layer in both saturated and non-saturated traffic
conditions taking into account the impact of both the physical propagation
channel and multirate transmission in Rayleigh fading environment.
The general traffic model assumed is M/M/1/K. Finally, it shows
that the behavior of the throughput in non-saturated traffic
conditions is a linear combination of two system parameters; the
payload size and the packet rates, $\lambda^{(s)}$, of each
contending station. The validity interval of the proposed model is
also derived.

Simulation results closely match the theoretical derivations,
confirming the effectiveness of the proposed models.
\end{abstract}
\begin{keywords}
DCF, Distributed Coordination Function, fading, IEEE 802.11, MAC,
Rayleigh fading, rate adaptation, saturation, throughput,
unsaturated, non-saturated.
\end{keywords}
%
%
\section{Introduction}
The IEEE802.11 Medium Access Control (MAC)
layer~\cite{standard_DCF_MAC} presents a mandatory option, namely
the Distributed Coordination Function (DCF), based on the Carrier Sense Multiple Access
Collision Avoidance CSMA/CA access method, that has received
considerable attention in the past years
\cite{Bianchi}-\cite{cantieni}.

Many papers, following the seminal work by Bianchi~\cite{Bianchi},
have addressed the problem of modeling the DCF in a variety of
traffic load and channel transmission conditions. Most of them
focus on a scenario presenting $N$ saturated stations that
transmit towards a common Access Point (AP) under the hypotheses
that the packet rates, along with the probability of transmission
in a randomly chosen time slot, is common to all the involved
stations, while the error events on the transmitted packets are
mainly due to collisions between packets belonging to different
stations.

Real networks are different in many respects. First, traffic is
mostly non-saturated, so it is important to derive a model
accounting for practical network operations. Channel conditions
are far from being ideal and often packet transmission has to be
rescheduled until data is correctly received. Due to Rayleigh and
shadow fading conditions, a real scenario presents stations
transmitting at different bit rates, because of multirate adaptation
foreseen at the physical layer of WLAN protocols such as IEEE
802.11b. Furthermore, other scenarios usually occur in a real
network. As an example, different stations usually operate with
different loads, i.e., they have different packet rates, while the
transmitting bit rate can also differ between the contending
stations. In all these situations the common hypothesis, widely
employed in the literature, that all the contending stations have
the same probability of transmitting in a randomly chosen time
slot, does not hold anymore. The aim of this paper is
to provide a model and theoretical analysis under much more
realistic scenarios. With this background, let us provide
a quick survey of the recent literature related to the problem
addressed in this paper.

In~\cite{Bianchi}, the author provides an analysis of the
saturation throughput of the basic 802.11 protocol assuming a two
dimensional Markov model at the MAC layer. Papers
\cite{HCheolLee}-\cite{Chatzimisios} model the influence of real
channel conditions on the throughput of the DCF operating in
saturated traffic conditions, while
\cite{zorzi_rao}-\cite{Spasenovski} thoroughly analyze the
influence of capture on the throughput of wireless transmission
systems.

The behavior of the DCF of IEEE 802.11 WLANs in unsaturated
traffic conditions has been analyzed in a number of papers
\cite{Liaw}-\cite{Daneshgaran_unsat}. In~\cite{Liaw} the authors
extend the Bianchi's model by introducing a new state in order to
consider non-saturated traffic conditions, accounting for the case
in which the station queue is empty after successful completion of
a packet transmission. In the modified model however, a packet is
discarded after $m$ backoff stages, while in Bianchi's model the
station keeps iterating in the $m$-th backoff stage until the
packet gets successfully transmitted. These models rely on the
basic hypotheses of ideal channel transmission and on the absence
of capture effects. Paper~\cite{Malone} proposes to model
non-saturated traffic conditions by adding a new state for each
backoff stage accounting for the absence of new packets to be
transmitted. In
\cite{Daneshgaran_unsat}-\cite{Daneshgaran_unsat_globecom}, the
authors extend the multi-dimensional Markovian state transition
model characterizing the MAC layer behavior by including
transmission states that account for packet transmission failures
due to errors caused by propagation through the channel, along
with a state characterizing the situation when a station has no
packets to transmit.

In \cite{Qiao}, the authors look at the impact of channel induced
errors and of the received SNR on the achievable throughput in a
system with rate adaptation, whereby the transmission rate of the
terminal is modified depending on either direct or indirect
measurements of the link quality. In \cite{heusse}, the authors
were the first to observe that in multirate networks the aggregate
throughput is strongly influenced by that of the slowest
contending station; such a phenomenon is denoted as ``performance
anomaly of the DCF of IEEE 802.11 protocol''. In \cite{Joshi},
authors provide an analytical framework for analyzing the link
delay of multirate networks. In \cite{cantieni}, authors provide
DCF models for finite load sources with multirate capabilities,
while in \cite{DuckYongYang} authors propose a DCF model for
networks with multirate stations and derive the saturation
throughput. Remedies to performance anomalies are also discussed.
In both previous works, packet errors are only due to collisions
between different contending stations.

In this paper, we substantially extend the previous work proposed in
the companion papers
\cite{Daneshgaran_unsat}-\cite{Daneshgaran_unsat_globecom} by
looking at all the three issues outlined before together, namely,
real channel conditions, saturated and non-saturated traffic, and
multirate capabilities. Our basic assumptions are essentially
similar to those of Bianchi~\cite{Bianchi}, with the exception
that we do assume the presence of both channel induced errors due
to the transmission over Rayleigh fading channel, and we consider
a general traffic model. As a reference standard, we use network
parameters belonging to the IEEE802.11b protocol, even though the
proposed mathematical model holds for any flavor of the IEEE802.11
family or other wireless protocols with similar MAC layer
functionality. We also derive a simple model of the aggregate
throughput in unloaded traffic conditions and show that
essentially, the throughput depends on the linear combination of
the packet rates $\lambda^{(s)}$ of each contending station, each
properly weighted by its payload size, while packet errors due to
imperfect channel conditions do not affect the aggregate
throughput. The validity interval of the proposed model is also
derived.

The paper is organized as follows. After a brief review of the
functionalities of the contention window procedure at MAC layer,
Section~\ref{Section_MarkovianModelCharacterizing} substantially
extends the Markov model initially proposed by Bianchi, presenting
modifications that account for transmission errors over Rayleigh
fading channels employing the 2-way handshaking technique in a
variety of traffic conditions.
Section~\ref{Section_MarkovianProcessAnalysis} is devoted to the
solution of the proposed bi-dimensional Markov chain related to
each contending station, and provides an expression for the
aggregate throughput of the link. After defining the basic time
parameters needed for system performance evaluation,
Section~\ref{Section_AverageTimeSlotDuration} estimates the
expected time slot duration needed for the evaluation of the
aggregate throughput of the link. The adopted traffic model is
discussed in Section~\ref{Section_TrafficModel}.

Section~\ref{PHY_section} briefly addresses the modeling of the
physical layer of IEEE 802.11b in a variety of channel conditions.
In Section~\ref{Simulation_results_section} we present simulation
results where typical MAC layer parameters for IEEE802.11b are
used to obtain throughput values as a function of various system
level parameters and SNR, under typical traffic conditions.
Section~\ref{Section_Alinearmodeloftheaggregate} derives a simple
model of the aggregate throughput in unsaturated traffic
conditions, along with its interval of validity. Insights on the
DCF behavior with multiple contending stations are also given,
considering a multirate setting. Finally,
Section~\ref{Section_Conclusions} is devoted to conclusions.
\section{Markovian Model Characterizing the MAC Layer under general traffic
conditions and Real Transmission Channel}
\label{Section_MarkovianModelCharacterizing}
In a previous paper~\cite{Daneshgaran_unsat}, we proposed a
bi-dimensional Markov model for characterizing the behavior of the
DCF under a variety of real traffic conditions, both non-saturated
and saturated, with packet queues of small sizes, and considered
the IEEE 802.11b protocol with the basic 2-way handshaking
mechanism. Many of the basic hypotheses were the same as those
adopted by Bianchi in the seminal paper~\cite{Bianchi}.

As a starting point for the derivations which follow, we adopt the
bi-dimensional model proposed in the companion
paper~\cite{Daneshgaran_unsat}, appropriately modified in order to
account for a scenario of $N$ contending stations, each one
employing a specific bit rate and a different transmission packet
rate. As in~\cite{Daneshgaran_unsat}, we consider the IEEE 802.11b
protocol employing the basic 2-way handshaking mechanism. For
conciseness, we refer the interested to~\cite{Daneshgaran_unsat}
for many details on the considered bi-dimensional Markov model.

Consider the following scenario: $N$ stations transmit towards a
common AP in infrastructure mode, whereby each station,
characterized by its own traffic load, can access the channel using
a data rate in the set $\{1, 2, 5.5, 11\}$ Mbps, depending on
channel conditions. Each bit rate is associated with a different
modulation format, whereas the basic rate is 1 Mbps with DBPSK
modulation (2Mbps with DQPSK if short preamble is used)
\cite{standard_DCF_MAC}. We identify a generic station with the
index $s \in S = \{1,2, \cdots, N\}$, where $N$ is the number of
stations in the network, and $S$ is the set of station indexes. As
far as the transmission data rate is concerned, we define four
rate-classes identified by a \textit{rate-class identifier} $r$
taking values in the set $R = \{1,2,3,4\}$ ordered by increasing values of
data rates $R_D = \{1,2,5.5,11\}$ Mbps (as an example, rate-class
$r = 3$ is related to the bit rate 5.5 Mbps). Concerning control
packets and PLCP header transmissions, the basic rate is
identified by $R_C$. Two basic rates are available for control
packets, namely 1 and 2 Mbps, whereby the latter is adopted for
short preambles \cite{standard_DCF_MAC}.

The traffic load of the s-th station is identified by a Packet
Arrival Rate (PAR) $\lambda^{(s)}$. Upon defining both
rate-classes and traffic, we can associate a generic station $s$
to a rate-class $r \in R$ and traffic load $\lambda^{(s)}$. As a
consequence, with respect to the model in
\cite{Daneshgaran_unsat}, we need to specify the probabilities along with
different Markov chains for each station in the network.

The two sources of errors on the transmitted packets are
collisions between packets and channel induced errors. Considering
the $s$-th station, collisions can occur with probability
$P_{col}^{(s)}$, while transmission errors due to imperfect
channel transmissions can occur with probability $P_e^{(s)}$.
Notice that $P_e^{(s)}$ depends upon the station rate-class $r$,
which in turn depends on the Signal-to-Noise Ratio (SNR) received
by the AP (appropriate expressions will be provided in Section
\ref{PHY_section} for each rate-class). We assume that collisions
and transmission error events are statistically independent. In
this scenario, a packet from the $s$-th station is successfully
transmitted if there is no collision (this event has probability
$1-P_{col}^{(s)}$) and the packet encounters no channel errors
during transmission (this event has probability $1-P_{e}^{(s)}$).
The probability of successful transmission is therefore equal to
$(1-P_{e}^{(s)})(1-P_{col}^{(s)})$, while the
probability of failed transmission is,
\begin{equation}\label{eq.equ}
P^{(s)}_{eq}=P^{(s)}_{col}+P^{(s)}_e-P^{(s)}_e\cdot  P^{(s)}_{col}
\end{equation}
To simplify the analysis, we make the assumption that the impact
of the channel induced errors on the packet headers are negligible
because of their short length compared to the data payload
size. This is justified on the basis of the assumption that the
bit errors inflicting the transmitted data are independent of each
other. Hence, the packet or frame error rate, identified
respectively with the acronyms PER or FER, is a function of the
packet length, with shorter packets having exponentially smaller
probability of error compared to longer packets. We note that with
sufficient interleaving we can always ensure that the errors
affecting individual bits in a data packet are independent of each
other.
\figuragrossa{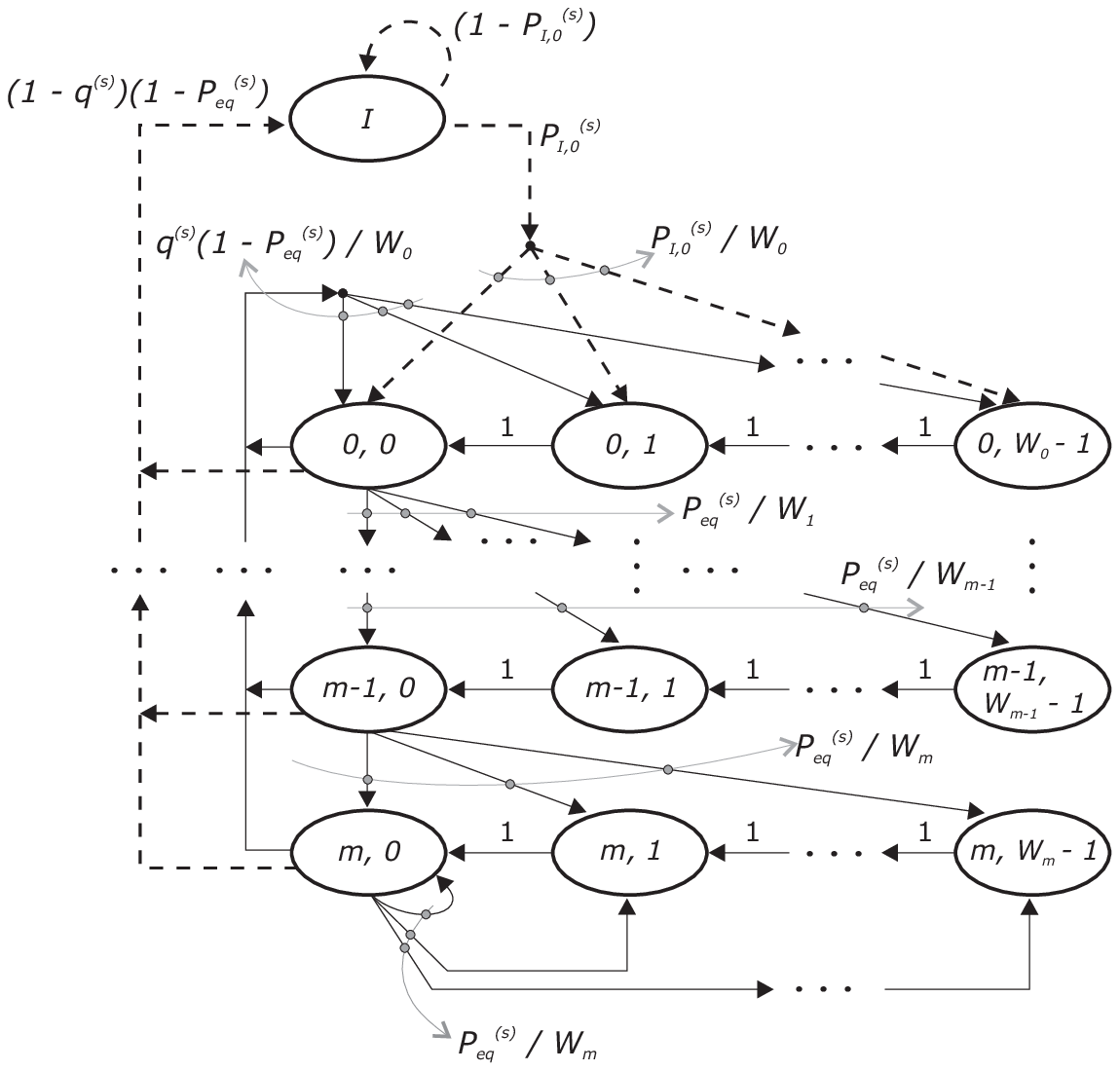}{Markov chain for the
contention model of the generic $s$-th station in general traffic
conditions, based on the 2-way handshaking technique, considering
the effects of channel induced errors.}{fig.chain}

The modified Markov model related to the $s$-th contending station
is depicted in Fig.~\ref{fig.chain}. We consider $(m+1)$ different
backoff stages, including the zero-th stage. The maximum
Contention Window (CW) size is $W_{max} = 2^mW_0$, and the
notation $W_i = 2^iW_0$ is used to define the $i^{th}$ contention
window size ($W_0$ is the minimum contention window size). A
packet transmission is attempted only in the $(i,0)$ states,
$\forall i=0,\ldots,m$. In case of collision, or due to the fact
that transmission is unsuccessful because of channel errors, the
backoff stage is incremented, so that the new state can be
$(i+1,k)$ with uniform probability $P_{eq}^{(s)}/W_{i+1}$. The
contention window is assumed to be common to all the stations in
the network; for this reason the station index $s$ is dropped from
the contention model depicted in Fig.~\ref{fig.chain}.

In order to account for non-saturated traffic conditions, we have
introduced a new state labelled $I$, accounting for the following
two situations:
\begin{itemize}
    \item Immediately after a successful transmission, the queue of the
    transmitting station is empty. This event occurs with
    probability $(1-q^{(s)})(1-P_{eq}^{(s)})$, whereby $q^{(s)}$ is the probability that there is at least one packet in the
    queue after a successful transmission (see Section~\ref{Section_TrafficModel}).
    \item The station is in an idle state with an empty queue until a new packet arrives in the
    queue. Probability $P_{I,0}^{(s)}$ represents the probability that
    while the station resides in the idle state $I$ there is at least one packet arrival, and a
    new backoff procedure is scheduled.
\end{itemize}
We note that the probability $P_{I,0}$ of staying in the idle
state is strictly related to the adopted traffic model. This issue
will be further investigated.

The transition probabilities\footnote{$P_{i,k|j,n}^{(s)}$ is short
for
$P^{(s)}\{s(t+1)^{(s)}=i,b(t+1)^{(s)}=k|s(t)^{(s)}=j,b(t)^{(s)}=n\}$.}
for the generic $s$-th station's Markov process in
Fig.~\ref{fig.chain} are as follow:
\begin{equation}\label{eq.process}\small
\begin{array}{lll}
P_{i,k|i,k+1}^{(s)} &= 1,                     &~ k \in [0,W_i-2], ~ i \in [0,m] \\
P_{0,k|i,0}^{(s)}  &= q^{(s)}(1-P_{eq}^{(s)})/W_0, &~ k \in [0,W_0-1], ~ i \in [0,m] \\
P_{i,k|i-1,0}^{(s)}   &= P_{eq}^{(s)}/W_i,&~ k \in [0,W_i-1], ~ i \in [1,m] \\
P_{m,k|m,0}^{(s)}   &= P_{eq}^{(s)}/W_m,&~ k \in [0,W_m-1]\\
P_{I|i,0}^{(s)}  &= (1-q^{(s)})(1-P_{eq}^{(s)}), &~ i \in [0,m]\\
P_{0,k|I}^{(s)}  &= P_{I,0}^{(s)}/W_0, &~ k \in [0,W_0-1] \\
P_{I|I}^{(s)}  &= 1-P_{I,0}^{(s)} &
\end{array}
\end{equation}
The first equation in~(\ref{eq.process}) states that at the
beginning of each backoff slot time, the backoff time is
decremented. The second equation accounts for the fact that after
a successful transmission, a new packet transmission starts with
backoff stage 0 with probability $q^{(s)}$, in case there is a new
packet in the queue to be transmitted. Third and fourth equations
deal with unsuccessful transmissions and the need to reschedule a
new contention stage. The fifth equation deals with the practical
situation in which after a successful transmission, the queue of
the station is empty, and as a consequence, the station transits
in the idle state $I$ waiting for a new packet arrival. The sixth
equation models the situation in which a new packet arrives in the
queue while the station is in the idle state. Finally, the seventh
equation models the situation in which there are no packets to be
transmitted and the station is in the idle state.
\section{Markovian Process Analysis}
\label{Section_MarkovianProcessAnalysis}
This section focuses on the evaluation of the stationary state distribution of the Markov model
proposed in the previous section. The objective is to find the probability that a station occupies
a given state at any discrete time slot:
\[
b_{i,k}^{(s)}=\lim_{t\rightarrow
\infty}P[s^{(s)}(t)=i,b^{(s)}(t)=k],~\forall
k\in[0,W_i-1],~\forall i\in[0,m]
\]
along with the stationary probability $b_I^{(s)}$ of being in the
idle state. This mathematical derivation is at the core of
finding the probability $\tau^{(s)}$ that a station will attempt
transmission in a randomly chosen time slot. In order to simplify
the notation, in what follows we will omit the superscript $(s)$ since
the mathematical derivations are valid for any contending station
$s=1,\ldots,N$.

For future developments, from the model depicted in
Fig.~\ref{fig.chain} we note the following relations:
\begin{equation}\label{trans_states_probabilities}
\begin{array}{rcll}
b_{i,0} & = & P_{eq} \cdot b_{i-1,0} = P_{eq}^i \cdot b_{0,0},   &\forall i \in [1,m-1] \\
b_{m,0} & = & \frac{P_{eq}^m}{1-P_{eq}} \cdot b_{0,0},           & i = m \\
\end{array}
\end{equation}
whereby, $P_{eq}$ is the equivalent probability of failed
transmission\footnote{For simplicity, we assume that at each
transmission attempt, any station will encounter a constant and
independent probability of failed transmission, $P_{eq}$,
independently from the number of retransmissions already suffered
by each station.} that takes into account the need for a new
contention due to either packet collision ($P_{col}$) or channel
errors ($P_e$).

Let us focus on the meaning of the idle state $I$ noted in
Fig.~\ref{fig.chain} to which the stationary probability $b_I$ is
associated. This state considers both the situation in which after
a successful transmission there are no packets in the station queue to be transmitted,
and the situation in which the packet queue
is empty and the station is waiting for new packet arrivals. The
stationary probability of being in state $b_I$ can be evaluated as
follows:
\begin{equation}
\label{eq:b_N}
\begin{array}{lll}
b_I & = & (1-q)(1-P_{eq})  \sum_{i=0}^{m}b_{i,0} + (1-P_{I,0})  b_I \\
    & = & \frac{(1-q)(1-P_{eq})}{P_{I,0}} \cdot
    \sum_{i=0}^{m}b_{i,0}
\end{array}
\end{equation}
Upon employing the probabilities $b_{i,0}$ noted
in~(\ref{trans_states_probabilities}), it is straightforward to
obtain:
\begin{equation}\label{eq:b00}
\sum_{i=0}^{m}b_{i,0} =b_{0,0}\left[
\sum_{i=0}^{m-1}P^i_{eq}+\frac{P_{eq}^m}{1-P_{eq}}\right]
=\frac{b_{0,0}}{1 - P_{eq}}
\end{equation}
By using the previous result,~(\ref{eq:b_N}) simplifies to
\begin{equation}\label{eq:bI}
b_I = \frac{1 - q}{P_{I,0}} \cdot b_{0,0}
\end{equation}
Equ.~(\ref{eq:b_N}) reflects the fact that state $b_I$ can be
reached after a successful packet transmission from any state
$b_{i,0},~\forall i \in [0,m]$ with probability $(1-q)(1-P_{eq})$,
or because the station remains in idle state with probability
$(1-P_{I,0})$, whereby $P_{I,0}$ is the probability of having at
least one packet to be transmitted in the queue. The statistical
model of $P_{I,0}$ will be discussed in
Section~\ref{Section_TrafficModel}. It is anticipated that
$P_{I,0}$ is related to the specific traffic model chosen for
modeling packet arrivals in the station queue.

The other stationary probabilities for any $k\in[1,W_i-1]$ follow
by resorting to the state transition diagram shown in
Fig.~\ref{fig.chain}:
\begin{equation}\label{eq.bik}
b_{i,k} = \frac{W_i-k}{W_i} \left\{
\begin{array}{ll}
q(1-P_{eq}) \cdot \sum_{i=0}^{m}b_{i,0} +       & \\
+ P_{I,0}
\cdot
b_I,& i = 0 \\
P_{eq} \cdot b_{i-1,0},                                      & i \in [1,m-1] \\
P_{eq} (b_{m-1,0} + b_{m,0}),                                & i = m \\
\end{array}\right.
\end{equation}
Upon substituting~(\ref{eq:b_N}) in~(\ref{eq.bik}), $b_{0,k}$ can
be obtained as follows:
\begin{equation}
\begin{array}{ll}
q(1-P_{eq}) \cdot \sum_{i=0}^{m}b_{i,0} + P_{I,0} \cdot b_I =&

\\
q(1-P_{eq}) \cdot \sum_{i=0}^{m}b_{i,0}+ P_{I,0} \cdot
\frac{(1-q)(1-P_{eq})}{P_{I,0}} \cdot \sum_{i=0}^{m}b_{i,0} =&
 \\
          = (1-P_{eq}) \cdot \sum_{i=0}^{m}b_{i,0}&
\end{array}
\end{equation}
Employing the normalization condition, after some mathematical
manipulations, and remembering~(\ref{eq:b00}), it is possible to
obtain:
\begin{eqnarray}
\label{eq:norm}
1 & = & \sum_{i=0}^{m}\sum_{k=0}^{W_i-1} b_{i,k} + b_I \nonumber \\
  & = & \frac{b_{0,0}}{2} \left\{ W_0\left[ \sum_{i=0}^{m-1}(2P_{eq})^i + \frac{(2P_{eq})^m}{1-P_{eq}} \right] + \frac{1}{1-P_{eq}}\right\} + b_I \nonumber \\
  & = & \alpha \cdot b_{0,0}+ b_I
\end{eqnarray}
whereby,
\begin{equation}
\label{eq:alpha} \alpha=\frac{1}{2} \left\{ W_0\left[
\frac{1-(2P_{eq})^m}{1-2P_{eq}} + \frac{(2P_{eq})^m}{1-P_{eq}}
\right] + \frac{1}{1-P_{eq}}\right\}
%
\end{equation}
From (\ref{eq:norm}), the following equation for computation of
$b_{0,0}$ easily follows:
\begin{equation}\label{eqb_00_norm}
b_{0,0} = \frac{1-b_I}{\alpha}
\end{equation}
Equ.~(\ref{eqb_00_norm}) is used to compute $\tau^{(s)}$, the
probability that the $s$-th station starts a transmission in a
randomly chosen time slot. In fact, taking into account that a
packet transmission occurs when the backoff counter reaches zero,
we have:
\begin{eqnarray}\label{eq:tau_s}
\tau^{(s)} &=  &  \sum_{i=0}^{m}b_{i,0}^{(s)}
=\frac{b_{0,0}^{(s)}}{1-P_{eq}^{(s)}}
=\frac{1-b_I^{(s)}}{\alpha^{(s)}(1-P_{eq}^{(s)})}= \\
&=  &
\frac{2(1-b_I^{(s)})(1-2P_{eq}^{(s)})}{(W_0+1)(1-2P_{eq}^{(s)}) +
W_0P_{eq}^{(s)}(1-(2P_{eq}^{(s)})^m)}\nonumber
\end{eqnarray}
whereby, we re-introduced the superscript $(s)$ since this expression will
be used in what follows.

The collision probability $P_{col}^{(s)}$ needed to compute
$\tau^{(s)}$ can be found considering that using a 2-way
hand-shaking mechanism, a packet from a transmitting station
encounters a collision if in a given time slot, at least one of
the remaining $(N-1)$ stations transmits one packet. Since each
station has its own $\tau^{(s)}$, the collision probability for
the $s$-th contending station depends on the transmission
probabilities of the remaining stations as follows:
\begin{equation}\label{eq:pcol_s}
P_{col}^{(s)} = 1-\prod_{\substack{j = 1 \\ j \neq s}}^{N} (1-\tau^{(j)})
\end{equation}
Given the set of $N$ equations (\ref{eq.equ}) and
(\ref{eq:tau_s}), a non-linear system of $2N$ equations can be
solved in order to determine the values of $\tau^{(s)}$ and
$P_{col}^{(s)}$ for any $s=1,\ldots,N$; this is the operating
point corresponding to the $N$ stations in the network needed in
order to determine the aggregate throughput of the network,
defined as the fraction of time the channel is used to
successfully transmit payload bits:
\begin{equation}\label{eq:throughput_aggr_s}
  S = \sum_{s=1}^{N} \frac{1}{T_{av}}P_s^{(s)} \cdot (1 - P_e^{(s)}) \cdot PL
\end{equation}
whereby, the summation is over the throughput related to the $N$
contending stations, $PL$ is the average payload size, and
$T_{av}$ is the expected time per slot defined in the following.

Probabilities involved in~(\ref{eq:throughput_aggr_s}) are as
follows: $P_e^{(s)}$ is the PER of the $s$-th station due to
imperfect channel transmissions and $P_s^{(s)}$ is the
probability of successful packet transmission from the $s$-th
station.

In the next section, we derive the mathematical relations defining
both $T_{av}$ and the probabilities involved
in~(\ref{eq:throughput_aggr_s}).
\section{Estimating the Average Time Slot Duration}
\label{Section_AverageTimeSlotDuration}
In order to proceed further, we need to evaluate the average time
that a station spends in any possible state, i.e., the average
time slot duration $T_{av}$, in terms of the key probabilities
involved in the proposed model. This is the focus of the current
section.

The average duration $T_{av}$ of a time slot (or expected time per
slot) can be evaluated by weighting the times spent by a station
in a particular state with the probability of being in that state.
It is possible to note four kinds of time slots.
\begin{itemize}
    \item The average idle slot duration, identified by $T_{I}$, in which no station is transmitting over the
    channel.

\item The average collision slot duration, identified by $T_{C}$,
in which more than one station is attempting to gain access to
the channel.

\item The average duration of the slot due to erroneous
transmissions because of imperfect channel conditions, identified
by $T_{E}$.

    \item The average slot duration of a successful transmission, identified by
    $T_{S}$.
\end{itemize}
\subsection{The average idle slot duration}
The average idle slot duration can be evaluated as the probability
$(1-P_t)$ that no station is attempting to gain access to the
channel times the duration $\sigma$ of an empty slot time.

Let $P_t$ be the probability that the channel is busy in a slot
because at least one station is transmitting. Then, we have:
\begin{equation}
P_t = 1 - \prod_{s=1}^{N} (1 - \tau^{(s)})
\end{equation}
The average idle slot duration can be defined as:
\begin{equation}
T_I = P_{I} \cdot \sigma= (1-P_{t}) \cdot \sigma
\end{equation}
where each idle slot is assumed to have duration $\sigma$.
\subsection{The average slot duration of a successful transmission}
Consider a tagged station in the set of $N$ stations in the
analyzed network, and let $s$ be its index in the set
$\{1,\ldots,N\}$. Then, the average slot duration of a successful
transmission can be found upon evaluating the probability that
only the $s$-th tagged station is successfully transmitting over
the channel, i.e.,
\begin{equation} \label{eq:ps_s}
P_s^{(s)}  =  \tau^{(s)} \prod_{\substack{j = 1 \\ j \neq s}}^{N}
(1 - \tau^{(j)})
\end{equation}
times the duration of a successful transmission from the $s$-th
station, $T_s^{(s)}$. The latter depends on the rate-class $(r)$
which the tagged station belong to, and can be evaluated as
follows:
\begin{eqnarray} \label{eq:TS_s}
T_s^{(s)} & = & \frac{H_{PHY}}{R_C} + \frac{H_{MAC} + PL}{R_D^{(s)}} + \delta + \\
          &   & +SIFS + \frac{H_{PHY} + ACK}{R_C} + \delta +
          DIFS\nonumber
\end{eqnarray}
whereby, $PL$ is the average payload length, $H_{PHY}$ and
$H_{MAC}$ are, respectively, the physical and MAC header sizes,
$\tau_p$ is the propagation delay, DIFS is the duration of the
Distributed InterFrame Space, $R_C$ is the basic data rate used
for transmitting protocol data, and $R_D^{(s)}$ is the data rate
of the $s$-th station.

With this setup, the average slot duration of a successful
transmission can be evaluated as:
\begin{equation} \label{eq:TS}
T_S = \sum_{i=1}^{N} P_s^{(i)}\left(1-P_e^{(i)}\right) \cdot
T_s^{(i)}
\end{equation}
whereby, $\left(1-P_e^{(i)}\right)$ accounts for the probability of
packet transmission without channel induced errors.
\subsection{The average collision slot duration}
In a network of stations transmitting equal length packets with
different data rates, the average duration of $T_{C}$ is largely
dominated by the slowest transmitting stations. This phenomenon is
called \textit{performance anomaly} of 802.11b, and it has been
first observed in \cite{heusse}.

As an example, suppose that a frame transmitted by a station using
the rate 1 Mbps (class 1) collides with the one of a station
transmitting at rate 11 Mbps one (class 4). Of course, both frames
get corrupted while the channel appears busy to the remaining
sensing stations for the whole duration of the frame transmitted
by the low rate station. This implies that fast stations (high
classes) are penalized by slow stations (low classes), causing a
drop of the aggregate throughput. The duration of the collision
depends on the frame duration of the lowest rate stations. In
order to evaluate the collision probability, we define the class
$(r)$ collision duration as follows:
\begin{equation}\label{eq:TC_r}
T_c^{(r)} = \frac{H_{PHY}}{R_C} + \frac{H_{MAC} + PL}{R_D^{(r)}} + ACK_{timeout}
\end{equation}
which takes into account basic rate $R_C$ and data rate $R_D^{(r)}$ of class $(r)$.

For the derivations which follow, we consider a set of indexes
which identify the stations transmitting with the $r$-th data
rate:
$$
n(r) = \{\mbox{identifiers of stations belonging to rate-class
(r)}\}
$$
$\forall r \in R=\{1,\ldots,4\}$ such that $\sum_{r=1}^{N_R}
|n^{(r)}| = N$ ($|\cdot|$ is the cardinality of the embraced set).

With this setup, there are two different kinds of collisions:
\begin{itemize}
\item intra-class collisions between \textit{at least} two frames
belonging to same rate class $(r)$;
\item inter-class collisions between \textit{at least} one frame
of class $(r)$ and \textit{at least} one frame of class $(j) >
(r)$
\end{itemize}
As far as intra-class $(r)$ collisions are concerned, the
collision probability $P_{c1}^{(r)}$ can be evaluated as follows:
\begin{eqnarray}\small \label{eq:pc1_r}
 \left\{ 1 - \left[\prod_{s \in n(r)} (1 - \tau^{(s)}) + \sum_{s \in n(r)} \tau^{(s)}
\prod_{\substack{j \in n(r) \\ j \neq s}}(1 - \tau^{(j)})\right] \right\} \cdot  &  &\nonumber \\
             \cdot \prod_{s \in \{S - n(r)\}} (1 - \tau^{(s)}) &   &
\end{eqnarray}
Notice that the latter is the probability that stations not
belonging to the same data rate set $n(r)$ do not transmit, times
the probability that there are at least two stations in the same
rate class $n(r)$ transmitting over the channel
\cite{DuckYongYang}. Notice that the first product within brace
brackets accounts for the scenario in which no stations with rate
in $n(r)$ transmit, or there is only one station transmitting with
rate in $n(r)$. As a side note, notice that $P_{c1}^{(r)}=0$ in
case there are no collisions between stations belonging to the
same rate class.

Following a similar reasoning, the inter-class $(r)$ collision
probability $P_{c2}^{(r)}$ can be evaluated as:
\begin{eqnarray} \label{eq:pc2_r}
\left[ 1 - \prod_{s \in n(r)} (1 - \tau^{(s)}) \right] \cdot
                   \left[ 1 - \prod_{j=r+1}^{N_R} \prod_{s \in n(j)} (1 - \tau^{(s)}) \right] \cdot &&\nonumber \\
       \cdot \left[ \prod_{j=1}^{r-1} \prod_{s \in n(j)} (1 - \tau^{(s)})
             \right]&&
\end{eqnarray}
which considers the scenario in which at least one station of
class $(r)$ and at least one station belonging to a higher rate
class $(j)$ (i.e., $(j)
> (r)$) transmit in the same time slot, while all the other stations belonging
to lower indexed classes (i.e., with $(i) < (r)$) are silent. As a
side note, notice that $P_{c2}^{(r)}=0$ in case there are no
collisions between stations belonging to different rate classes.

The total class $(r)$ collision probability is the sum of the
previous two probabilities: 
\begin{equation} \label{eq:pc_r}
P_c^{(r)} = P_{c1}^{(r)} + P_{c2}^{(r)}
\end{equation}
while the average collision slot duration can be computed
considering the whole set of classes $r \in R$ and their collision
probabilities weighted by their durations:
\begin{equation} \label{eq:TC}
T_C = \sum_{r=1}^{N_R} P_c^{(r)} \cdot T_c^{(r)}
\end{equation}
\subsection{The average duration of the slot due to erroneous
transmissions}
The average duration of the slot due to erroneous transmissions
can be evaluated in a way similar to the one used for evaluating
$T_S$ and $T_C$:
\begin{equation} \label{eq:TE}
T_E = \sum_{i=1}^{N} P_s^{(i)} \cdot P_e^{(i)} \cdot T_e^{(i)}
\end{equation}
whereby, $P_s^{(i)}$ is defined in (\ref{eq:ps_s}), and $T_e^{(s)}$
is assumed to be equal to $T_c^{(s)}$ since when a channel error
occurs, the transmitting station does not receive the
acknowledgment before the end of the ACK timeout.
\subsection{Average time slot duration}
Given the expected slots derived in the previous sections, the
average duration of a slot time is simply:
\begin{equation} \label{eq:TAV}
T_{av} = T_I + T_C + T_S + T_E
\end{equation}
\section{Traffic Model}
\label{Section_TrafficModel}
This section presents the traffic model employed in our setup
along with the derivation of the key probabilities $q^{(s)}$ and
$P^{(s)}_{I,0}$ shown in Fig.~\ref{fig.chain}. The offered load
related to each station is characterized by the parameter
$\lambda^{(s)}$ representing the rate at which packets arrive at
the $s$-th station buffer from the upper layers, and measured in
packets per second. The time between two packet arrivals is
defined as \textit{interarrival time}, and its mean value is
evaluated as $\frac{1}{\lambda^{(s)}}$. One of the most commonly
used traffic models assumes that the packet arrival process is
Poisson. The resulting interarrival times are exponentially
distributed.

In the proposed model shown in Fig.~\ref{fig.chain}, we need to
know the probability $q^{(s)}$ that there is at least one packet
to be transmitted in the queue. Probability $q^{(s)}$ can be well
approximated in a situation with small buffer size \cite{Malone}
through the following relation:
\begin{equation}
\label{q} q^{(s)} = 1 - e^{- \lambda^{(s)} T_{av}}
\end{equation}
where, $T_{av}$ is the \textit{expected time per slot}, which is
useful to relate the state of the Markov chain with the actual
time spent in each state. Such a time has been derived in
(\ref{eq:TAV}). Under the hypothesis of small queue systems,
probabilities $q^{(s)}$ and $P_{I,0}^{(s)}$ can be evaluated
considering that the probability of having at least one packet
arrival in the queue at the end of a successful packet
transmission can be approximated with the probability of having at
least one packet arrival in an average time slot duration. As a
result of this simple approximation, we have $q^{(s)} =
P_{I,0}^{(s)}$. Upon remembering~(\ref{eq:bI})
and~(\ref{eqb_00_norm}), $\tau ^{(s)}$ in~(\ref{eq:tau_s}) can be
evaluated as:
\begin{eqnarray} \label{eq:tau_s_smallQ}
\tau ^{(s)}  =  \frac{q^{(s)}\left(1 -
P_{eq}^{(s)}\right)^{-1}}{q^{(s)} (\alpha^{(s)} - 1) + 1}
=\frac{2(1-2P_{eq}^{(s)})q^{(s)}}{D(q^{(s)},W_0,m,P_{eq}^{(s)})}&&
\end{eqnarray}
whereby,
\begin{equation}\small\label{D_q_pe}
\begin{array}{ll}
D(q^{(s)},W_0,m,P_{eq}^{(s)})=q^{(s)}\left[(W_0+1)(1-2P_{eq}^{(s)})
+
W_0P_{eq}^{(s)}\cdot\right.&\\
\left.(1-(2P_{eq}^{(s)})^m)\right] +2(1-q^{(s)})(1-P_{eq}^{(s)})(1-2P_{eq}^{(s)})&\\
\end{array}
\end{equation}
Though simple, this approximation has been verified by simulation,
proving to be quite effective for predicting the aggregate
throughput.
\section{Physical Layer Modeling}
\label{PHY_section}
In a scenario with $N$ contending stations randomly distributed
around a common access point, throughput performance depends on
the channel conditions experienced by each station.
In what follows we briefly recall the main signal propagation
issues in order to evaluate the PER experienced by a generic
station in the network. This will serve as the basis for the
simulated scenarios discussed in
Section~\ref{Simulation_results_section}.
\subsection{Signal Propagation}
Consider a contending station at distance $d$ from an access
point. Given the one-sided noise power spectral
density\footnote{In what follows we will set the effective antenna
temperature $T=273K$ and $N_o=-174$dBm.}, $N_o$,
%
%
the received SNR can be evaluated as~\cite{Rappaport}:
\begin{equation}\label{receiv_snr}
SNR_{\textrm{dB}}=P(d)\left|_{\textrm{dBm}}\right.-N_o-B_w\left|_{\textrm{dB}}\right.-N_F
\end{equation}
whereby, $N_F$ is the receiver noise figure (10dB), while
the power received at a
distance $d$ denoted $P(d)\left|_{\textrm{dBm}}\right.$ is given by,
\begin{equation}\label{receiv_power_d}
P(d)\left|_{dBm}\right.=P_{tx}\left|_{dBm}\right.-L(d)\left|_{dB}\right.
\end{equation}
In the previous equation, $L(d)\left|_{dB}\right.$ is the
path-loss at distance $d$ from the transmitter
\[
L(d)\left|_{dB}\right.=L_o\left|_{dB}\right.+10\cdot
n_p\log_{10}\left(\frac{d}{d_0}\right)
\]
and $L_o$ is defined as,
\[
L_o\left|_{dB}\right.=-10\log_{10}\left(\frac{G_tG_r
\lambda^2}{(4\pi)^2d_0^{n_p}}\right)
\]
where, $d_0$ is a reference distance (usually selected equal to 1m)
with path-loss $L_o$.

The other symbols are defined as follows; $G_t$ is the transmitter
antenna gain, $G_r$ is the receiver antenna gain, $n_p$ is the
path-loss exponent, and $\lambda=c/f$ is the wavelength given by
the ratio between the light velocity and the carrier frequency.
The path-loss exponent $n_p$ depends on the specific propagation
environment and it ranges from 2 (free space propagation) to
3.5-4 for non-line-of-sight propagation, or multi-path fast fading
conditions in indoor environments~\cite{Rappaport}. Furthermore,
based on FCC regulations, in the 2.4GHz ISM band the transmitted
power $P_{tx}\left|_{dBm}\right.$ is $20$dBm or
equivalently, $100$mW.

The SNR per transmitted bit $\gamma$ accounts for the spreading
gain $C_s/B_s$, and is defined as:
\begin{equation}\label{receiv_snr_per_bit}
\gamma\left|_{dB}\right.=SNR_{dB}+10\log_{10}\left(\frac{C_s}{B_s}\right)
\end{equation}
whereby, $C_s$ is the number of chips per symbol while $B_s$ is
the number of bits per transmitted symbol (see Table
\ref{80211b_setup}).
%
%
\begin{table}
\caption{PHY setup of the IEEE 802.11b standard.}
\begin{center}
\begin{tabular}{l|c|c|c|c}\hline
\hline

Frequency [GHz]& 2.4 &2.4 &2.4& 2.4\\

Bit rate [Mbps] &1 &2 &5.5& 11\\

Modulation& DBSPK & DQPSK & CCK &CCK\\

Chips per symbol, $C_s$ & 11 & 11 & 8 &8\\

Bits per symbol, $B_s$ & 1 & 2 & 4 &8\\

%
%

Channel band [MHz] &22 &22 &22& 22\\

%
%
%

Minimum power of received& -85 &-82&
-80& -76\\

 signal (Sensitivity) [dBm] &&&\\


\hline\hline
\end{tabular}
 \label{80211b_setup}
\end{center}
\end{table}
%
\subsection{IEEE 802.11b Physical Layer}
The physical layer (PHY) of the basic 802.11b standard is based on
the spread spectrum technology. Two options are specified, the
Frequency Hopped Spread Spectrum (FHSS) and the Direct
Sequence Spread Spectrum (DSSS). The FHSS uses Frequency Shift Keying
(FSK) while the DSSS uses Differential Phase Shift Keying (DPSK)
or Complementary Code Keying (CCK).

DSSS transmits signals in the 2.4-GHz ISM band (i.e.,
2.4000–-2.4835 GHz). The basic 802.11 DSSS gives data rates of 1
and 2 Mbps. The 802.11b extension \cite{standard_DCF_MAC} employs
DSSS at various rates including one employing CCK encoding on 4
and 8 bits for each CCK symbol, or optionally employing packet
binary convolutional coding. The four supported data rates in
802.11b are 1, 2, 5.5 and 11 Mbps.

BER performance of the various transmitting modes of IEEE802.11b
are shown in Table \ref{ber_various_conditions} for various
channel models \cite{standard_DCF_MAC_2,SimonAlouini,Fainberg}. In
Table \ref{ber_various_conditions}, the SNR in
(\ref{receiv_snr_per_bit}) is denoted as $\gamma$ while $I_k(ab)$
is the modified Bessel function of $k$th order.
\begin{table*}\caption{BER $P_b(\gamma)$ for various channel models}
\begin{center}
\begin{tabular}{c|l|l|l}\hline
\hline Channel Model & DBPSK & DQPSK & CCK-5.5/11 Mbps \\\hline
AWGN & $\frac{1}{2}\textrm{erfc}\left(\sqrt{\gamma}\right)$ &
$Q_1(a,b)-\frac{1}{2}I_0(ab)e^{-\frac{1}{2}a^2-\frac{1}{2}b^2}$ &
$1-\int_{-\sqrt{\gamma}}^{+\infty}\left[\int_{-(z+\sqrt{\gamma})}^{+(z+\sqrt{\gamma})}e^{-\frac{\eta^2}{2}}d\eta\right]^{\frac{\alpha}{2}-1}e^{-\frac{z^2}{2}}dz$\\

&&
$a=\sqrt{2\gamma\left(1-\sqrt{\frac{1}{2}}\right)},~b=\sqrt{2\gamma\left(1+\sqrt{\frac{1}{2}}\right)}$
& $\alpha=\left \{
\begin{array}{ll}
4, & \textrm{5.5 Mbps}\\
8, & \textrm{11 Mbps}
\end{array} \right.$
\\

&&$Q_1(a,b)=e^{-\frac{a^2+b^2}{2}}\sum_{k=0}^{+\infty}\left(\frac{a}{b}\right)^k I_k(ab)$&\\

&&$
I_0\left(ab\right)=1+\sum_{k=1}^{+\infty}\left[\frac{\left(ab/2\right)^k}{k!}\right]^2
$ &\\\hline

Rayleigh fading & $\frac{1}{2\left(1+\gamma\right)}$
&$\frac{1}{2}\left[1-\sqrt{\frac{\gamma
\frac{\sqrt{2}}{2}}{1+\gamma \frac{\sqrt{2}}{2}}}\right]$&
$\frac{2^{(\alpha-1)}}{2^{\alpha}-1}\sum_{i=1}^{\alpha-1}\frac{(-1)^{i+1}C^{\alpha-1}_i}{1+i+i\cdot \gamma}$\\

&&&$\alpha=\left \{
\begin{array}{ll}
4, & \textrm{5.5 Mbps}\\
8, & \textrm{11 Mbps}
\end{array} \right.$\\
&&& $C^{\alpha-1}_i=\frac{(\alpha-1)!}{i!\cdot (\alpha-1-i)!}$\\
%
%
\hline\hline
\end{tabular}
 \label{ber_various_conditions}
\end{center}
\end{table*}
The FER as a function of the SNR can be computed as follows:
\begin{equation}\label{fer_1}\small
P_e(\textrm{DATA},\textrm{SNR})=1-\left[1-P_e(\textrm{PLCP},\textrm{SNR})\right]\cdot
\left[1-P_e(\textrm{PSDU},\textrm{SNR})\right]
\end{equation}
where,
\begin{equation}\label{fer_2}\small
P_e(\textrm{PLCP},\textrm{SNR})=1-\left[1-P_b(\textrm{DBPSK},\textrm{SNR})\right]^{8\times
\textrm{PLCP}},
\end{equation}
and
\begin{equation}\label{fer_3}\small
P_e(\textrm{PSDU},\textrm{SNR})=1-\left[1-P_b(\textrm{TYPE},\textrm{SNR})\right]^{8\times
\textrm{PSDU}}.
\end{equation}
In the previous equations, $\textrm{PLCP}$ is the sum of the lengths
of the $\textrm{PLCP}$ Preamble and the $\textrm{PLCP}$
header, $\textrm{PSDU}$ is the sum of the lengths of the payload size
$\textrm{PL}$ and the $\textrm{MAC}$ header and
$\textrm{DATA}=\textrm{PSDU}+\textrm{PLCP}$.

$P_b(\textrm{DBPSK},\textrm{SNR})$ is the BER as a function of SNR
for the lowest data transmit rate employing DBPSK modulation (or
DQPSK if the basic rate 2Mbps is adopted). Note that the FER,
$P_e(\textrm{DATA},\textrm{SNR})$, implicitly depends on the
modulation format used. Hence, for each supported rate, one curve
for $P_e(\textrm{DATA},\textrm{SNR})$ as a function of SNR can be
generated. $P_b(\textrm{TYPE},\textrm{SNR})$ is modulation
dependent whereby the parameter $\textrm{TYPE}$ can be any of the
following $\textrm{TYPE}\in
\{\textrm{DBPSK},\textrm{DQPSK},\textrm{CCK5.5},\textrm{CCK11}\}$\footnote{The
acronyms are short for Differential Binary Phase Shift Keying,
Differential Quadrature Phase Shift Keying and Complementary Code
Keying with rate $5.5$ or $11$ Mbps, respectively.}.
\section{Simulation Results and Model Validation}
\label{Simulation_results_section}
\begin{table}\caption{Typical network parameters}
\begin{center}
\begin{tabular}{c|c||c|c}\hline
\hline MAC header & 28 bytes&Propag. delay $\tau_p$ & 1 $\mu s$\\
\hline PLCP Preamble & 144 bit & PLCP Header & 48 bit \\
\hline PHY header & 24 bytes&Slot time & 20 $\mu s$\\
\hline PLCP rate & 1Mbps & W$_0$ & 32\\
\hline No. back-off stages, m & 5 & W$_{max}$ & 1024\\
\hline Payload size & 1028 bytes&SIFS & 10 $\mu s$\\
\hline ACK & 14 bytes&DIFS & 50 $\mu s$\\
\hline ACK timeout & 364$\mu s$ &EIFS & 364 $\mu s$\\

\hline\hline
\end{tabular}
 \label{tab.design.times}
\end{center}
\end{table}
This section focuses on simulation results for validating the
theoretical models and derivations presented in the previous
sections. We have developed a C++ simulator modeling both the DCF
protocol details in 802.11b and the backoff procedures of a
specific number of independent transmitting stations. The
simulator considers an Infrastructure BSS (Basic Service Set) with
an AP and a certain number of fixed stations which communicate
only with the AP. For the sake of simplicity, inside each station
there are only three fundamental working levels; traffic model
generator, MAC and PHY layers. Traffic is generated following the
exponential distribution for the packet interarrival times.
Moreover, the MAC layer is managed by a state machine which
follows the main directives specified in the
standard~\cite{standard_DCF_MAC}, namely waiting times (DIFS,
SIFS, EIFS), post-backoff, backoff, basic and RTS/CTS access modes.
The typical MAC layer parameters for IEEE802.11b noted in
Table~\ref{tab.design.times}~\cite{standard_DCF_MAC} have been
used for performance validation.

For conciseness, in this paper we present a set of results related
to the three scenarios described in the following.

The first investigated scenario considers $N=10$ contending
stations. Nine stations are randomly placed on the perimeter of a circle of
radius $R$, while the AP is placed at the center of the
transmission area. Upon employing equations
(\ref{receiv_snr})-(\ref{receiv_snr_per_bit}) with $n_p=4$
(typical of heavily faded Rayleigh channel conditions), we have
chosen a distance $R=20$~m in such a way that the SNR per
transmitted bit is above the minimum sensitivity specified in
Table \ref{80211b_setup}, relative to $11$ Mbps bit rate. Such
stations are in saturated conditions and have a packet rate
$\lambda=8$ kpkt/s. The payload size, assumed to be common to all
the transmitting stations, is $PL=1028$ bytes. The tenth station
in the following is identified as the slow station and is placed at 4
different distances from the AP so that its transmission occurs
with four different bit rates envisaged in the IEEE802.11b protocol.

The theoretical aggregate throughput achieved by the system in this
scenario is depicted in Fig. \ref{PL1028N10_1slowstat} as a
function of the packet rate of the slow station. Curves in both
subplots have been parameterized with respect to the bit rate of
the slow station. Simulated points are denoted with cross-points
over the respective theoretical curves. The upper curves refer to
ideal channel conditions, i.e., PER=0, while the lower subplot
represents a scenario in which the packets transmitted by all the
stations are affected by a PER equal to $8\cdot 10^{-2}$, which is
the worst-case situation with regards to the minimum sensitivity
\cite{standard_DCF_MAC}.

Some considerations are in order. Both subplots show that the
aggregate throughput is significantly lower than 11 Mbps even
though all the stations transmit at the highest bit rate
(continuous curve). This is essentially due to overhead and
control data transmitted at the basic rate $R_C$. Moreover,
throughput reduces as the PAR $\lambda_{slow}$ increases,
reaching saturation values strongly influenced by the rate of the
slowest station. This phenomenon is called \textit{performance
anomaly} of 802.11b, and it has been first observed in
\cite{heusse}. A comparative analysis of the set of curves
depicted in both subplots reveals that a small throughput reduction
is due to the presence of channel induced errors.
\figura{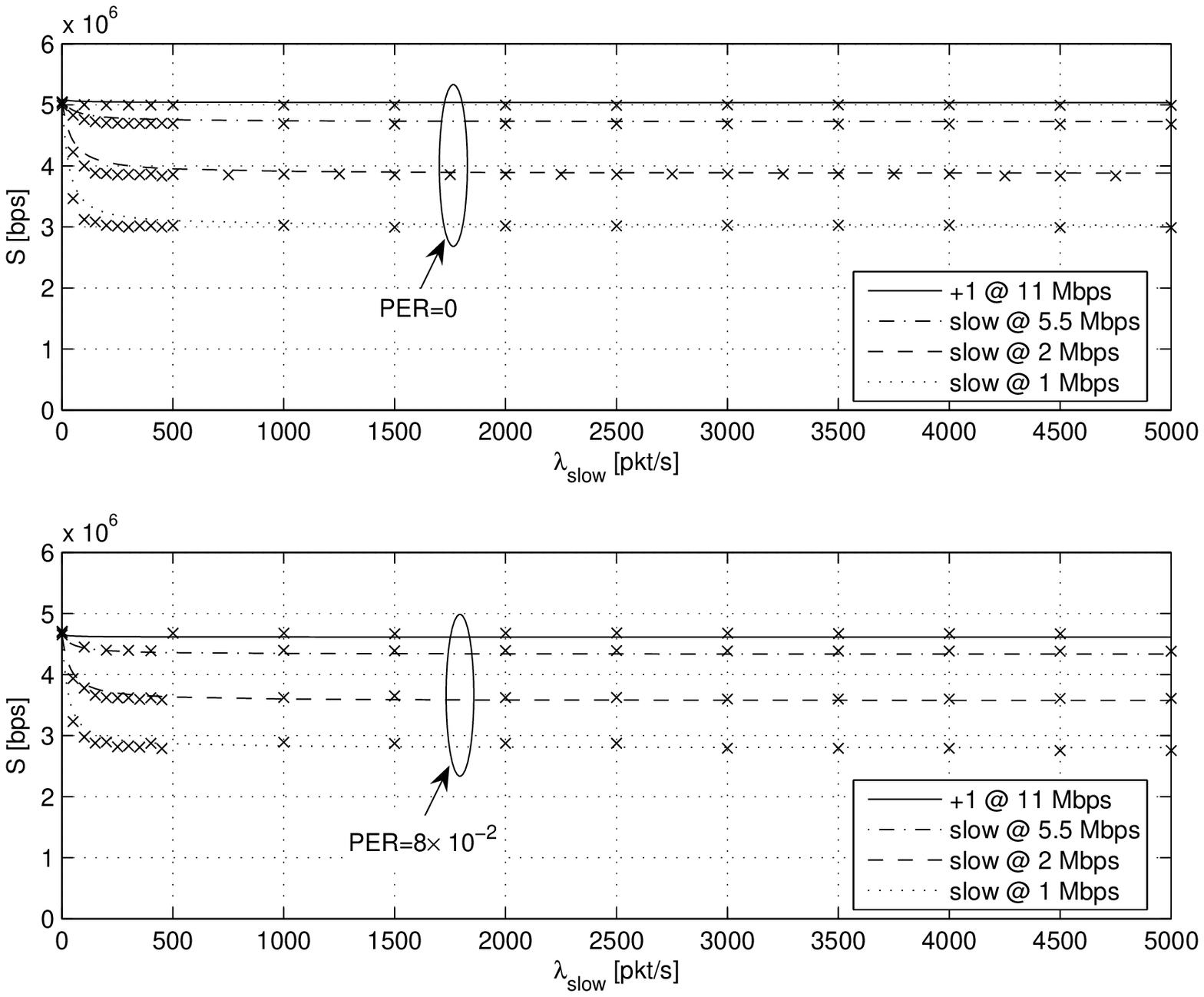}{Theoretical and simulated
throughput for the 2-way mechanism as a function of the packet
rate $\lambda_{slow}$ of the slow station, for four different bit
rates shown in the legends, and different PER values. Simulated
points are identified by cross-markers over the respective
theoretical curves. Payload size is 1028 bytes for all the $N=10$
contending stations, whereby 9 stations are saturated and transmit
with a packet rate of 8 kpkt/s at the maximum bit rate of $11$
Mbps. In the upper subplot $\textrm{PER}=0$, while in the lower
one $\textrm{PER}=8\times 10^{-2}$.}{PL1028N10_1slowstat}

The second investigated scenario is as follows. The four bit rates
are equally distributed between $8$ contending stations ($2$
stations with each bit rate). The packet rate $\lambda$ is the
same for all the contending stations and represents the
independent variable against which the aggregate throughput in
Fig.~\ref{PL1028N8_equi_distributed} is drawn. Curves have been
parameterized with respect to the PER identified in the legends,
assumed to be equal for all the contending stations. PER equal to
zero represents ideal channel transmission conditions, while
PER$=8\times 10^{-2}$ is the PER specified in the standard
\cite{standard_DCF_MAC}, corresponding to the minimum receiver
sensitivity.

In addition to noting the expected throughput reduction due to the packet
error rate, notice that the throughput manifests a linear behavior
for low values of the packet rates with a slope depending mainly
on the packet size of the contending stations. This observation
will be further analyzed in the next section where a linear model
for the throughput will be developed in unsaturated conditions.
Moreover, note that in unsaturated traffic conditions, the
aggregate throughput is quite independent of the packet errors
due to non ideal channel conditions.
\figura{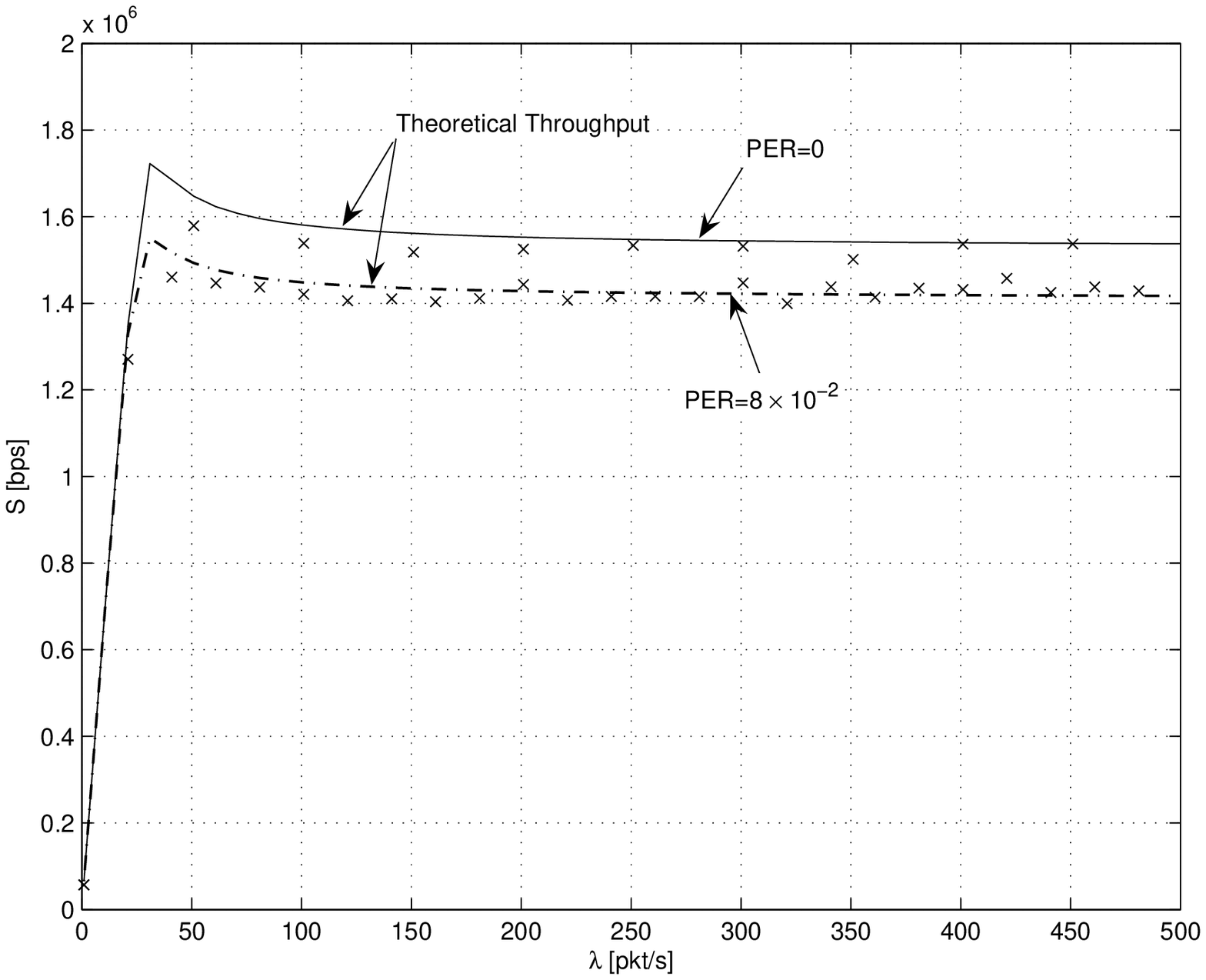}{Theoretical and simulated
throughput for the 2-way mechanism as a function of the packet
rate $\lambda$ of the 8 contending stations, organized so that the
four bit rates are equally distributed between the involved
stations. Simulated points are identified by cross-markers over
the respective theoretical curves. Payload size is
$\textrm{PL}=1028$ bytes for all the contending
stations.}{PL1028N8_equi_distributed}
%
%

The last investigated scenario considers three saturated stations
transmitting over a Rayleigh fading channel. Two stations
transmit at the maximum bit rate of 11 Mbps, since they are located
at 5 m distance from the AP. The third station is assumed to move radially
from the AP, and to switch between the four envisaged bit rates as
its distance from the AP increases. The aggregate throughput is
shown in Fig. \ref{Sat_throughput_3stations} as a function of the
distance of the third station from the AP. While the theoretical
throughput estimated with the proposed model is shown as a
continuous curve, simulated results are noted with star-marked
points. The radial distances from the AP at which rate switching
occurs are noted with vertical lines in the same plot. Rate
switching has been accomplished upon evaluating the minimum
distance at which the per-station PER (according to the
appropriate BER relation summarized in Table
\ref{ber_various_conditions} along with expressions
(\ref{fer_1})-(\ref{fer_3})) was above $8\times 10^{-2}$ for each
employed bit rate, as specified in the standard
\cite{standard_DCF_MAC,standard_DCF_MAC_2}.

As already noted in the results presented in the previous plots,
the aggregate throughput is strongly influenced by the bit rate of
the slowest station. Up to about 26 m, the 3 stations transmit at
the maximum bit rate and the aggregate throughput slightly
decreases because of the worst BER conditions due to the SNR
reduction of the moving station. As the distance increases and
the bit rate of the moving stations decreases, the behavior
changes. In fact, at higher bit rates, the larger the distance,
the more the moving station tends to go through backoff stages
because of the increasingly worse channel conditions. The channel
inactivity of the moving station is in favor of the other two
fastest stations and the aggregate throughput tends to slightly
increase.
\figura{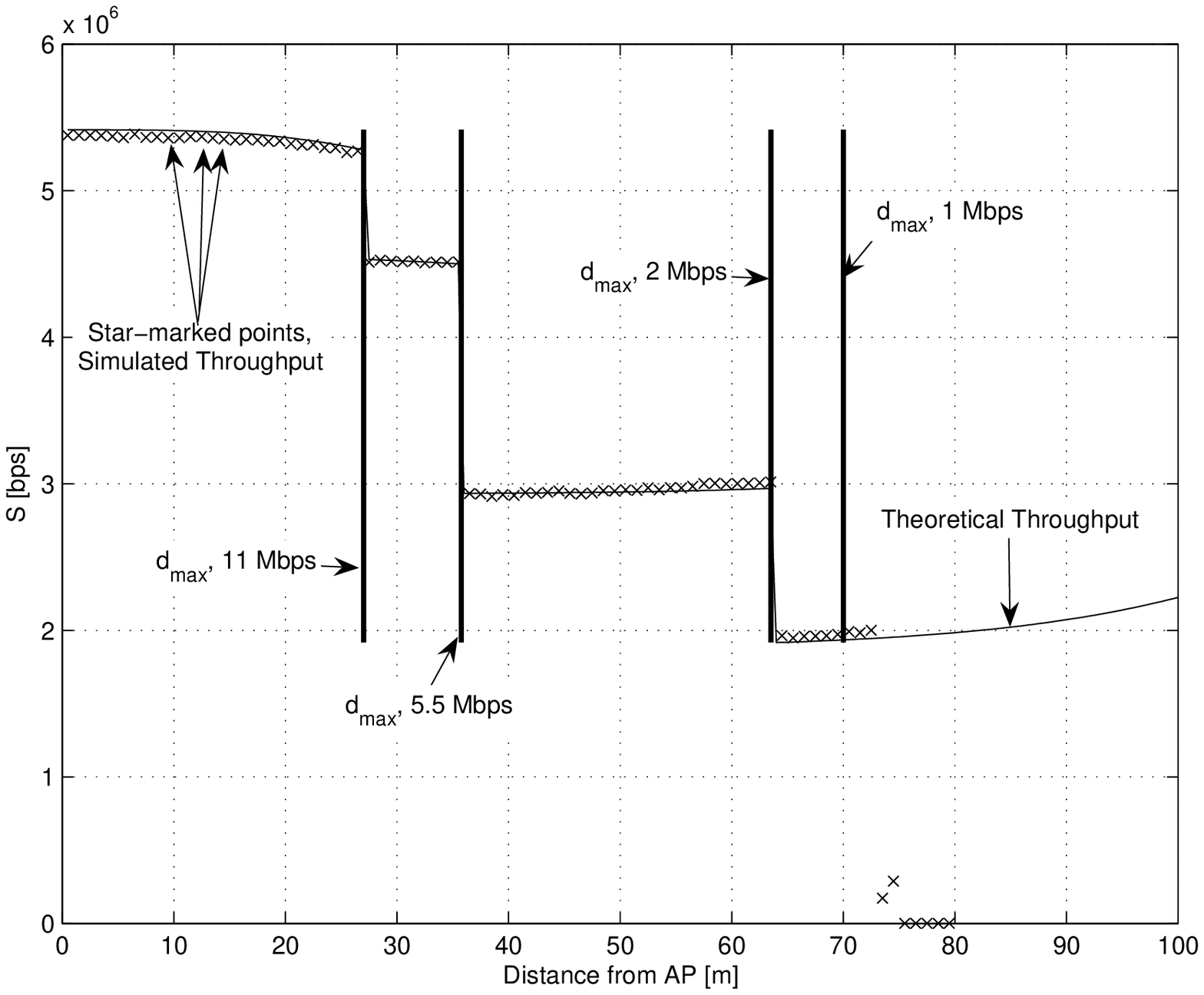}{Theoretical (continuous
curve) and simulated (star-marked points) throughput for the 2-way
mechanism for three saturated stations as a function of the
distance from the AP of the third station. Payload size is
$\textrm{PL}=1028$ bytes for all the contending
stations.}{Sat_throughput_3stations}
\section{A simple model of the aggregate throughput in non-saturated conditions}
\label{Section_Alinearmodeloftheaggregate}
This section derives a simple model of the aggregate throughput in
unsaturated conditions. To this end, consider the aggregate
throughput in~(\ref{eq:throughput_aggr_s})
\begin{equation}\label{eq:throughput_aggr_s_2}
S\left(\tau^{(1)},\ldots,\tau^{(N)}\right)=\sum_{s=1}^{N}
\frac{1}{T_{av}}P_s^{(s)} \cdot (1 - P_e^{(s)}) \cdot PL
\end{equation}
whereby, we have emphasized the dependence of
$S\left(\tau^{(1)},\ldots,\tau^{(N)}\right)$ on the set of $N$
probabilities $\tau^{(1)},\ldots,\tau^{(N)}$, with $\tau^{(s)}$
defined in~(\ref{eq:tau_s}). Of course, probabilities
$\tau^{(1)},\ldots,\tau^{(N)}$ depend on the traffic rates
$\lambda^{(1)},\ldots,\lambda^{(N)}$ of the $N$ contending
stations as exemplified by~(\ref{q}) and~(\ref{eq:tau_s_smallQ}).

Let us find an expression for
$S\left(\tau^{(1)},\ldots,\tau^{(N)}\right)$ when the overall
system approaches unsaturated traffic conditions. In this respect,
let us find the expressions for $P_s^{(s)}$, $P_e^{(s)}$ and
$T_{av}$ in the limit $\overline{\tau}\rightarrow \overline{0}$,
whereby, the previous compact relation is used to signify the fact
that all the probabilities $\tau^{(1)},\ldots,\tau^{(N)}$
approaches very small values.

First of all, notice that from~(\ref{q})
and~(\ref{eq:tau_s_smallQ}), the following approximations easily
follow:
\begin{equation}\label{eq:throughput_aggr_s_3}
\begin{array}{ll}
q^{(s)}\approx \lambda^{(s)} T_{av}, & \lambda^{(s)}\rightarrow 0,~ \forall s=1,\ldots,N\\
\tau^{(s)}\approx \frac{\lambda^{(s)} T_{av}}{1-P_{eq}^{(s)}}, &
\lambda^{(s)}\rightarrow 0,~ \forall s=1,\ldots,N
\end{array}
\end{equation}
From~(\ref{eq:pcol_s}), it easily follows
\[
\lim_{\overline{\tau}\rightarrow \overline{0}}P_{col}^{(s)}=0,~
\forall s=1,\ldots,N
\]
Given the relation
$P_{eq}^{(s)}=P_e^{(s)}+P_{col}^{(s)}-P_e^{(s)}P_{col}^{(s)}$ in
(\ref{eq.equ}), as $\overline{\lambda}\rightarrow \overline{0}$
(which is tantamount to considering $\overline{\tau}\rightarrow
\overline{0}$ in light of the relation
(\ref{eq:throughput_aggr_s_3})) we have:
\[
P_{eq}^{(s)}\rightarrow P_e^{(s)},~ \forall s=1,\ldots,N
\]
From~(\ref{eq:ps_s}), as $\overline{\tau} \rightarrow
\overline{0}$ the following relation holds,
\figura{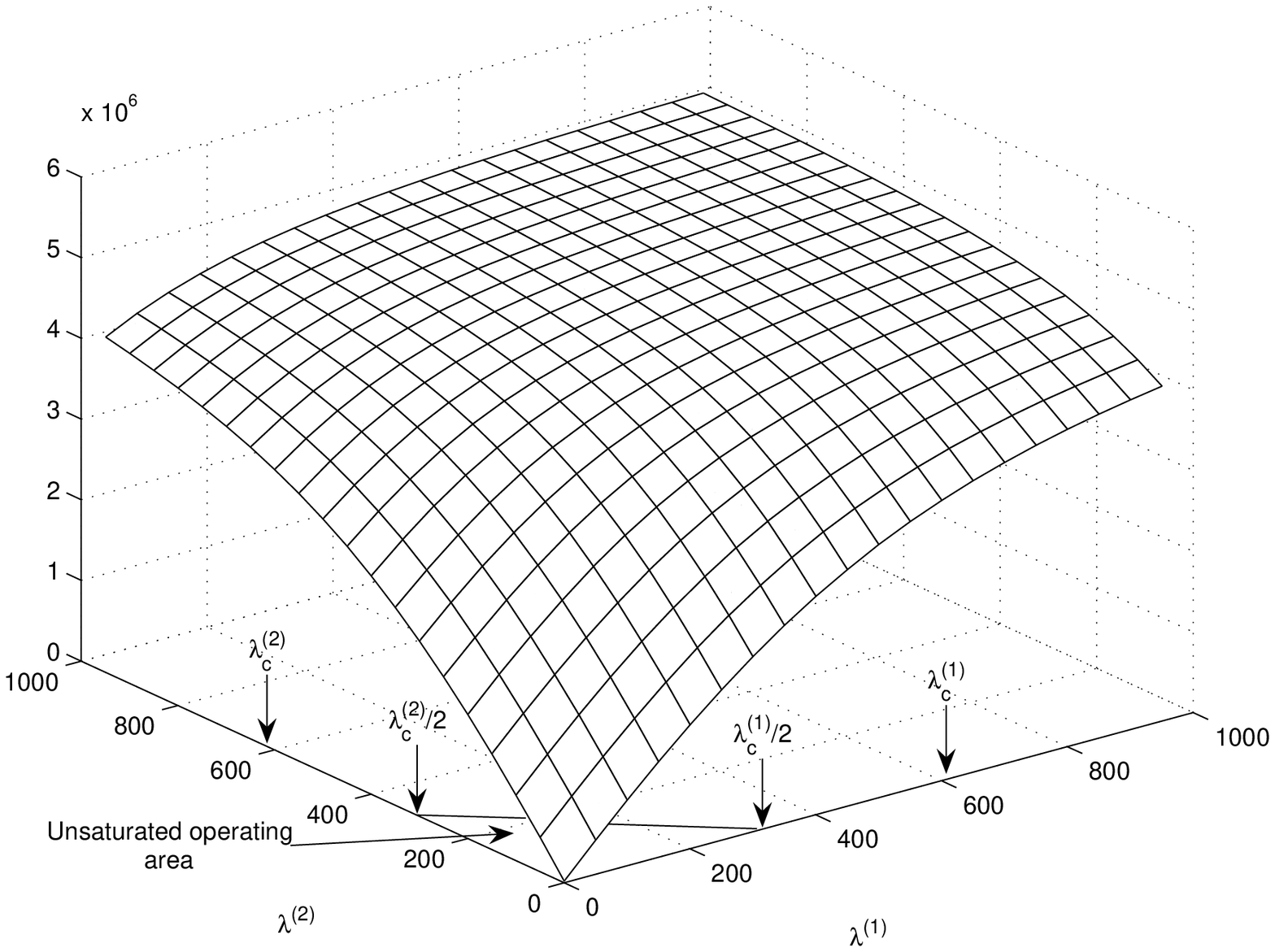}{Aggregate throughput for $N=2$, 11Mbps
contending stations over the bi-dimensional set $\lambda^{(1)}\in
[0,1000]~\textrm{pkt/s},\lambda^{(2)}\in [0,1000]~\textrm{pkt/s}$,
for $PL=1028$ bytes. The other transmission parameters are noted
in Table~\ref{tab.design.times}.}{results_opt_lin_model}
\[
P_s^{(s)}\approx \tau^{(s)}\approx \frac{\lambda^{(s)}
T_{av}}{1-P_{e}^{(s)}},~ \forall s=1,\ldots,N
\]
Upon employing the previous
relations,~(\ref{eq:throughput_aggr_s_2}) can be rewritten as:
\begin{equation}\label{eq:throughput_aggr_s_4}
S\left(\lambda^{(1)},\ldots,\lambda^{(N)}\right)\approx PL\cdot
\sum_{s=1}^{N} \lambda^{(s)}
\end{equation}
which is valid under the hypothesis that the packet length $PL$ is
the same for all the $N$ contending stations. In case of different
packet lengths, the aggregate throughput is the weighted sum
of each station packet rate times the respective packet
length.

Notice that in unsaturated conditions, the aggregate throughput is
independent from the packet error rate affecting data transmission
from each contending station in the network.

Equ.~(\ref{eq:throughput_aggr_s_4}) states that for very low
values of $\lambda^{(s)},~ \forall s=1,\ldots,N,$ the aggregate
throughput behaves as a linear combination of the packet rates of
the $N$ contending stations times the average payload length $PL$.
As a reference example, consider the aggregate
throughput\footnote{The throughput has been plotted using
(\ref{eq:throughput_aggr_s}) evaluated by employing the solution
of the non-linear system of $2N$ equations discussed at the end of
Section \ref{Section_MarkovianProcessAnalysis}.} shown in
Fig.~\ref{results_opt_lin_model} for a scenario with $N=2$
(11Mbps) contending stations. In this reference example, we
have considered ideal channel conditions (i.e., $P_e^{(s)}=0,~ \forall
s=1,2$) so that only collisions can yield packet losses. The other
transmission parameters are noted in Table~\ref{tab.design.times}.
There are various points that are worth noting from this figure.
When both stations approach saturated conditions, i.e., for
$\lambda^{(1)}$ and $\lambda^{(2)}$ tending to infinity, the
maximum aggregate throughput is less than half the maximum bit
rate of each contending station due to overhead and control data
transmitted at the basic rate $R_C$. On the other hand, when both
$\lambda^{(1)}$ and $\lambda^{(2)}$ take on very small values, the
aggregate throughput points define a plane predicted by
(\ref{eq:throughput_aggr_s_4}) in the region
$(\lambda^{(1)},\lambda^{(2)})\in D_2$ defined as follows
\begin{equation}\label{region_D}
D_2=\left\{
\begin{array}{ll}
\lambda^{(1)}\in \left[0,\frac{\lambda_c^{(1)}}{2}\right) &\\
0\le \lambda^{(2)}\le
-\frac{\lambda_c^{(2)}}{\lambda_c^{(1)}}\lambda^{(1)}+\frac{\lambda_c^{(2)}}{2}&
\end{array}
\right.
\end{equation}
whereby, $\lambda_c^{(1)}$ and $\lambda_c^{(2)}$ marked on
Fig.~\ref{results_opt_lin_model}, represents the limiting values
of the packet rate above which each station approaches saturated
traffic conditions\footnote{For the considerations that follow,
let us suppose the values of $\lambda_c^{(1)}$ and
$\lambda_c^{(2)}$ are known. Their values will actually be derived
later.}. Region $D_2$ along with the line\footnote{This
is the equation of the line passing through the two points
$(0,\lambda_c^{(2)}/2)$ and $(\lambda_c^{(1)}/2,0)$.}
$\lambda^{(2)}=-\frac{\lambda_c^{(2)}}{\lambda_c^{(1)}}\lambda^{(1)}+\frac{\lambda_c^{(2)}}{2}$
is identified as \textit{unsaturated operating area} in
Fig.~\ref{results_opt_lin_model}.

Consider now the previous scenario with two 11Mbps contending
stations transmitting at a fixed packet rate
$\lambda^{(1)}=\lambda^{(2)}=50$~\textrm{pkt/s} (unsaturated
conditions for both stations), and a third station starts
transmitting at a bit rate equal to 1 Mbps. As above, we
have considered ideal channel conditions for the two 11 Mbps stations
(i.e., $P_e^{(s)}=0,~ \forall s=1,2$) while the third station
is assumed to experience three different PER values $P_e^{(3)}=0, 10^{-1},
10^{-2}$.

Fig.~\ref{linearity_throughput_lambda} shows the aggregate
throughput as a function of the packet rate $\lambda^{(3)}$ of the
third station. It can be observed that when $\lambda^{(3)}$ takes
on low values, the overall system can be considered in unsaturated
conditions, and therefore model~(\ref{eq:throughput_aggr_s_4})
holds. In this case,
$S(\lambda^{(1)},\lambda^{(2)},\lambda^{(3)})$ can be considered
as a line with slope equal to the average packet length $PL$, while
$PL\left(\lambda^{(1)}+\lambda^{(2)}\right)=PL\cdot 100$ pkt/s is
the value of $S(\lambda^{(1)},\lambda^{(2)},\lambda^{(3)})$ for
$\lambda^{(3)}=0$. So long as $\lambda^{(3)}\in
[0,\lambda_c^{(3)}/2)$, model~(\ref{eq:throughput_aggr_s_4})
predicts the aggregate throughput quite well without the necessity of solving the
non-linear system described at the end of
Section~\ref{Section_MarkovianProcessAnalysis}. On the other hand,
when $\lambda^{(3)}$ grows above $\lambda_c^{(3)}/2$, the
system approaches saturated conditions quite fast and the overall
throughput drops asymptotically to $1.3$Mbps, as shown in
Fig.~\ref{linearity_throughput_lambda} for
$\lambda^{(3)}\rightarrow \infty$. Notice that the aggregate
throughput is strongly influenced by that of the slowest
contending station (1 Mbps in this specific case), confirming the
rate anomaly problem already noted in literature by
resorting to simulations \cite{heusse}. Also note that the PER
experienced by the third station as already demonstrated theoretically
above, does not affect the linear zone
of the aggregate throughput, while it reduces the aggregate throughput obtained in
saturated conditions. Also shown in the same figure is the
throughput of the third station (when the first two stations are
silent, i.e., $\lambda^{(1)}=\lambda^{(2)}=0$~\textrm{pkt/s})
along with the linear model of the throughput valid for values of
$\lambda^{(3)}$ lower than $\lambda_c^{(3)}$.
\figura{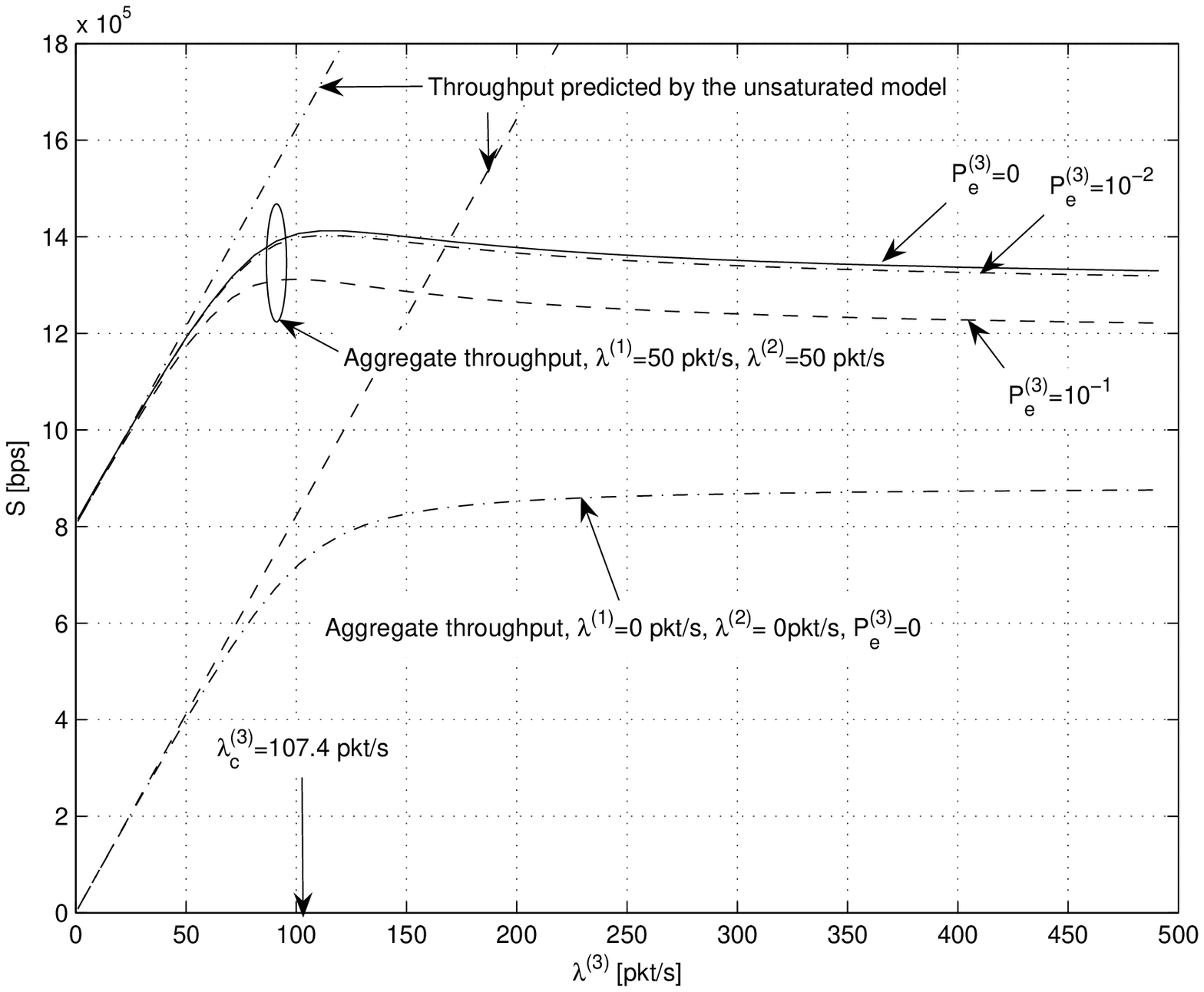}{Theoretical and
Simulated Throughput for a scenario with three contending stations
with the following transmission parameters: the first two stations
transmit with a bit rate of 11Mbps, while the third one is a 1Mbps
station. Payload size is 1028 bytes for all stations. Packet rates
are noted in the figure for the two scenarios investigated.
Dash-dotted line refers to the linear model of the throughput
derived
in~(\ref{eq:throughput_aggr_s_4}).}{linearity_throughput_lambda}

For $N$ contending stations, the region $D_N$ is the volume
contained in the region $(\lambda^{(1)},\ldots,\lambda^{(N)})\in
D_N$ defined as
\begin{equation}\label{region_D_N}
D_N=\left\{
\begin{array}{ll}
\lambda^{(1)}\in \left[0,\frac{\lambda_c^{(1)}}{2}\right) &\\
\ldots &\\
\lambda^{(N-1)}\in \left[0,\frac{\lambda_c^{(N-1)}}{2}\right) &\\
0\le \lambda^{(N)}\le
\frac{\lambda_c^{(N)}}{2}-\lambda_c^{(N)}\sum_{s=1}^{N-1}\frac{\lambda^{(s)}}{\lambda_c^{(s)}}&
\end{array}
\right.
\end{equation}
whereby, the last equation is the hyperplane passing through the
$N$-dimensional points
$$(\lambda_c^{(1)}/2,0,\ldots,0),(0,\lambda_c^{(2)}/2,\ldots,0),\ldots,(0,\ldots,0,\lambda_c^{(N)}/2)$$
The natural question at this point is the determination of the
critical packet rates $\lambda_c^{(s)},~ \forall s=1,\ldots,N,$
needed in order to identify the interval of validity of the
aggregate throughput in~(\ref{eq:throughput_aggr_s_4}). Consider
the $s$-th station in the network. Such a station is in saturation
when there is always a packet to be transmitted in its buffer. A
simple threshold on $\lambda^{(s)}$ discerning unsaturated from
saturated operation is $\lambda_c^{(s)}=\frac{1}{T_{st}}$,
whereby $T_{st}$ is the service time of the $s$-th station. The
reason is simple; the service time is the time interval from when
a packet is taken from the queue head, to when it is successfully
transmitted. If the packet inter-arrival time in the buffer is
greater than $T_{st}$, then on the average, the buffer will contain
at most one packet waiting for transmission. On the other hand,
when the packet inter-arrival time is less than $T_{st}$, then
packets will be buffered for successive transmissions. In this
respect, $T_{st}$ is the equilibrium interval based on which on
average a packet is transmitted soon after its arrival in the
buffer.

The service time $T_{st}$ can be defined as
follows~\cite{Liaw,WLee}:
\[
T_{st} = T_{A} + T_{TX}
\]
whereby, $T_A$ is the average time that a station spends through
the various backoff stages before transmitting a packet, i.e., the
so called MAC access time, while $T_{TX}$ is the average duration
of a transmission.

As far as $T_{A}$ is concerned, we note that despite the
existence of $m+1$ backoff stages (see Fig. \ref{fig.chain}), in
unsaturated conditions where the collisions between stations
rarely occur, a packet is transmitted in the $0$-th backoff stage
after an average number of backoff slots equal to $W_0/ 2$. Based
on these considerations, the average MAC access time in unloaded
traffic conditions can be simply defined as
\begin{equation}\label{equaz_T_A}
T_{A} =\frac{W_0}{2}\sigma
\end{equation}
whereby, $\sigma$ is the duration of an empty slot.

The average duration of a transmission, $T_{TX}$, can be evaluated
by observing that a station can experience two possible
situations; successful transmission or collision. Upon neglecting
the effects of collisions based on the considerations stated above in
connection to $T_A$, $T_{TX}$ can be defined as in
(\ref{eq:TS_s}); $T_{TX}=T_s^{(s)}$.

The values of $\lambda_c^{(r)}$ derived under these hypotheses,
are shown in Table~\ref{tab.lambdacritici} as a function of the
station transmission bit rate.
\begin{table}\caption{Critical Packet Rates.}
\begin{center}
\begin{tabular}{l||c|c|c|c}\hline\hline

Station bit rate & 1 Mbps & 2 Mbps & 5.5 Mbps & 11 Mbps \\\hline

$\lambda_c^{(r)}$-[pkt/s]&    107.4 & 196.5 & 416.3 & 612.0\\

\hline\hline
\end{tabular}
 \label{tab.lambdacritici}
\end{center}
\end{table}
\section{Conclusions}
\label{Section_Conclusions}
In this paper, we have presented a multi-dimensional Markovian
state transition model characterizing the DCF behavior at the MAC
layer of the IEEE802.11 series of standards by accounting for
channel induced errors and multirate transmission typical of
fading environments, under both non-saturated and saturated
traffic conditions. The model provided in the paper allows taking into consideration
the impact of channel contention in throughput analysis which is
often not considered or is considered with a static model by
using a mean contention period. Subsequently, based on justifiable
assumptions, the stationary probability of the Markov chain is
calculated obtaining the throughput in both non-saturated and
saturated conditions.

Finally, we have shown that the behavior of the aggregate throughput
in non-saturated traffic conditions is a linear combination of the
payload size along with the packet rates of each contending
station in the network. This result is significant for system level optimization
of the network. Theoretical derivations were supported by simulations.
\end{document}